# Hidden role of metastable phases on surface tension and in the selection of solid polymorphs from melt

**Puja Banerjee and Biman Bagchi***

*Solid State and Structural Chemistry Unit, Indian Institute of Science, Bangalore-560012, India*

___________________________________________________________________

## *Abstract*

*The preferential formation of one solid over the other, as it precipitates out from the melt at specific temperatures, is often explained by invoking a competition between thermodynamic and kinetic control. A quantitative theory, however, could not be developed because of the lack of accurate values of relevant surface tension terms. Motivated by the observations that wetting of the interface between two stable phases by multiple metastable phases of intermediate order can reduce the surface tension significantly (Kirkpatrick-Thirumalai-Wolynes (KTW), Phys. Rev. A 1989, 40 (2), 1045; Santra et al., J. Phys. Chem. B, 2013, 117, 13154 ), we develop a statistical mechanical approach based on a Landau-Ginzburg type free energy functional to calculate the surface tension between two stable phases in the presence of N number of metastable phases. Simple model calculations are performed that show the surface tension between two coexisting stable phases (melt and the stable crystalline forms) depends significantly on the number, relative depths and arrangements of the free energy minima of the metastable phases, in addition to the size of the nucleus. We provide an explanation of the quickly disappearing polymorphs (QDPMs) that often melt back to the liquid (or, the sol) phase. It is shown that our model systems could describe some aspects of solid formation in real polymorphic systems, like phosphates and zeolites.*

Email to: profbiman@gmail.com

# I. Introduction

The wetting the interface between two stable phases (melt and stable solid) by multiple intermediate metastable phases is predicted to lower the surface tension significantly[1-3,4]. In such situations, the surface tension between melt and stable solid (SS) is predicted to be significantly lower. It was further predicted that the surface tension could depend on the radius of the nucleus, thus lowering the nucleation free energy barrier and enhancing the rate of nucleation. Classical nucleation theory (CNT) can include such effects through an experimental surface tension, although not the radius dependence. Also, CNT cannot describe the trapping of those metastable intermediate phases in any satisfactory manner [5]. Using classical density functional theory (DFT) or computer simulation technique one can show that the extent of metastability of intermediate states plays an important role to introduce non-classical pathway of nucleation and control polymorph selection [6,7]. Among many limitations of CNT, the one dimensional approach with only the free energy gap and the surface tension at coexistence stand out to be the most glaring one.

The concept of thermodynamically versus kinetically controlled structure formation has found wide use in understanding the paradoxes presented by the formation of many biologically and industrially important crystals. Zeolites present a good example. Depending on the Si/Al ratio and hydration, it forms many polymorphs with different density and stability as faujasite (FAU), mordenite (MOR), stilbite (STI), quartz[8]. Although quartz is the stable thermodynamic phase and faujasite is one of the least stable metastable phases, quartz precipitates out from the melt sodium alumino silicate only at high temperature while faujasite forms at low temperature. At intermediate temperatures one can detect the formation of other zeolite phases that of intermediate stability or metastability.

Zeolite is not the only example. Phosphate and titanate series of solids also demonstrates such apparent anomalous selection of phases. In a recent experimental study, metal phosphate, LiFePO$_4$(olivine structure) has been shown to exhibit many metastable intermediate phases while crystallization from amorphous melt to the stable crystal forms[9] and many more. Titanates also form complex solids, with Magnasium Titanate being a prominent example. Other than these, silica (SiO$_2$) forms many polymorphs depending on thermodynamic and kinetic stability such as α-quartz, β-quartz, tridymite, cristobalite, coesite and stishovite[8, 10, 11, 12]. These forms are interconvertible by changing temperature, pressure etc. Another classic example of polymorphism is calcium carbonate which can exist in three different forms in nature: calcite, aragonite, vaterite[13].

Yet another type of nucleation is found to occur in the system of zirconia (ZrO$_2$) at an ambient pressure though the following pathway

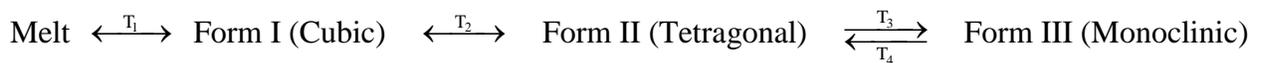

Melt $\xleftrightarrow{T_1}$ Form I (Cubic) $\xleftrightarrow{T_2}$ Form II (Tetragonal) $\xrightleftharpoons[T_4]{T_3}$ Form III (Monoclinic)

Here tetragonal form to monoclinic form transformation occurs though a diffusionless, military transformation, that is called martensite transformation[14-15]. In this nucleation, atoms move less than one atomic spacing. Other than ZrO$_2$, this type of transformation plays an important role in hardening of alloys such as steel, many other iron alloys, Ni-Ti etc.[16] In these systems, a high temperature face centred cubic lattice (fcc) phase is quenched to a body centred cubic (bcc) or body centred tetragonal (bct) phase. This involves a complete different mechanism and cannot be explained by theories based on CNT.

One encounters several such examples in pharmaceutical industry where one often uses the term "disappearing polymorphs" to characterize the metastable phases with transient existence [17]. We shall discuss it later, in detail. Appearance of metastable phases is explained by invoking the logic of kinetic stability. This rather odd combination of terms implies that the metastable phases are selected because of their lower nucleation barrier compared to the stable phase. Somehow the discussion has remained semi-quantitative. Nevertheless, the study of nucleation for these systems with multiple polymorphs has received attention from both theory and experimental studies for last few decades[1-3, 6, 18]. The final form of the crystals precipitates out from a melt bulk phase often observed to be influenced by the kinetic pathway of phase separation and the aspects of kinetic versus thermodynamic control of crystal nucleation has been studied extensively in past[19, 20]. The phrase "kinetically controlled" signifies that the metastable phases which have lowest nucleation barrier forms in a faster rate than others, whereas the formation of most stable solid forms by "thermodynamic control" which is favorable at higher temperature as it helps to overcome the larger nucleation barrier, often associated to the most stable phase.

In 1897, Ostwald described the significance of metastable phases in between supersaturated melt phase and stable crystal[21]. According to this, at a lower temperature, the least stable state forms first, then the second least and so on. Therefore, the ultimate control of a crystallization process of a polymorphic system lies in the choice of intermediate metastable phases in the phase transformation kinetics. Ostwald's step rule (OSR) states that the phase that would form first from the melt is the one that bears closest similarity to the melt. This rule has been used to explain the nucleation of polymorphic crystalline systems such as silica, carbonates, phosphates, iron oxides, sulfides, zeolite (aluminosilicate) etc[21]. The theoretical understanding of OSR is based on the classical nucleation theory (CNT) that predicts a nucleation free energy barrier as a combination of surface tension, $\gamma$ and free energy difference, $\Delta G_v$ between the melt (the parent phase) and the new, precipitating phase.

However, lack of the estimates of thermodynamic data, especially of the surface tension between two phases, have hindered progress. Note that experimental measurement of surface tension is hard as it requires the coexistence of two phases. For example, we would need faujasite and quartz to coexist to obtain the surface tension between the two phases. The scenario is a little better for free energy estimates and some have appeared in the literature.

The objectives of our present work are the following:

i) Classical nucleation theory fails to treat the effects of metastable states on phase transformations in many different scenarios. We discuss transient phase trapping and the role of surface tension and free energy of the metastable phase in the trapping. The failure of CNT lies in the one order parameter description. The importance of multidimensional nucleation theory has been discussed. We also present an analysis of the lifetime of disappearing polymorph. These are done in section II.

ii) We want to analyze the change in surface tension between two stable phases in the presence of N number of metastable phases. To this aim, we have modeled two kinds of polymorphic systems: one in which all the metastable phases have comparable free energy minima and the other with cascade geometry of free energy minima of metastable phases. To calculate surface tension between two stable phases we have considered them to be at coexistence for both the model systems.

iii) In section III and IV we use Cahn-Hilliard theory to calculate surface tension between two stable phases in the absence of any metastable phases and compare it to that in the presence of N number of metastable phases. Within the Cahn-Hilliard theory, surface tension decreases with the number of metastable phases and exhibits a dependence on the location and curvature of free energy surfaces.

iv) Section V presents our numerical calculations of surface tension in two model systems. Our results demonstrate the change in surface tension between two stable phases as a function of number of intermediate metastable phases for different characteristics of free energy surfaces and their energetics.

v) Polymorphism can be controlled by changing temperature and pressure of the system. In section VI, we have analyzed the change in surface tension between melt and solid phase as temperature is changed.

vi) In section VII, we have shown some important polymorphic system that can be understood using our models. Section VIII contains a discussion on multidimensional barrier surface and in section IX we conclude our work.

## II. Prediction of Classical Nucleation Theory (CNT) on selection of polymorphs

CNT focuses entirely on free energy considerations along the traditional reaction coordinate (which is the size of the critical nucleus). Here we apply classical nucleation theory (CNT) to understand the sequential phase transformation of metastable phases to stable solid phase. According to CNT, the free energy barrier of nucleation is defined as

$$\Delta G(R) = -\frac{4\pi}{3} R^3 \Delta G_v + 4\pi R^2 \gamma \tag{1}$$

This is known as the "capillarity approximation" which derives contributions from two opposite terms, one with energy gain, and the other with energy loss. Surface energy between two phases causes the increase in the free energy barrier of nucleation where $\Delta G_v$ (difference in free energy per unit volume) contributes in the opposite way and in the competition of these two, the free energy of nucleation shows a maximum value for a particular nucleus size, that is called the critical nucleus. By maximizing the functional form (Eq. (1)), one obtains the critical nucleus, R* and the free energy barrier at the critical cluster size ($\Delta G^*$)

$$R^* = \frac{2\gamma}{\Delta G_v}$$

$$\Delta G^* = \frac{16\pi}{3} \frac{\gamma^3}{(\Delta G_v)^2} \tag{2}$$

Therefore, the nucleation barrier could be more sensitive to the surface tension term than the free energy gap. This is essentially the rational for Ostwald step rule: closeness lowers surface tension, and in turn, the free energy barrier even though the free energy gap dictates the opposite. This could also serve as the essence of kinetic versus thermodynamic control.

Let us make the above discussion quantitative. For nucleation of stable solid (SS) phase in melt, there should be a free energy stabilization ( $\Delta G_{Melt/SS} < 0$ ). But, for the nucleation of metastable state (MS) from melt, $\Delta G_{Melt/MS} > 0$. Therefore, the free energy of nucleation for the latter will increase monotonically with nucleus size (R), where the nucleation free energy of the previous one shows a maxima at critical nucleus (R*) and then decreases. However, if we think their nucleation free energy curves cross at some value of nucleus size ( $R'$ ), we can write

$$-\frac{4\pi}{3}R'^3\Delta G_{Melt/SS} + 4\pi R'^2\gamma_{Melt/SS} = \frac{4\pi}{3}R'^3\Delta G_{Melt/MS} + 4\pi R'^2\gamma_{Melt/MS}$$

$$R' = \frac{3(\gamma_{Melt/SS} - \gamma_{Melt/MS})}{(\Delta G_{Melt/SS} + \Delta G_{Melt/MS})} \quad (3)$$

Now, if $\gamma_{Melt/SS} > \gamma_{Melt/MS}$, $R'$ is positive and the two nucleation is sequential. On the other hand, according to CNT, the critical nucleus for the nucleation of stable solid (SS) is

$$R^* = \frac{2\gamma_{Melt/SS}}{\Delta G_{Melt/SS}} \quad (4)$$

Therefore, the individual values of surface tension and free energy difference decides the critical nucleus size of MS from which SS can grow( $R' > R^* \, or \, R' < R^*$) and the effective nucleation barrier.

Now, let us assume the model of melt phase and stable bulk phase (in coexistence as per the requirement of surface tension calculation) in the presence of three metastable phases, all of them has higher free energy than the two stable phases (**Figure 1(a)**). Now, we will analyze the free energy profile of each phase as well as the size of nuclei for each phase using CNT.

Here we assume that each new phase grows from the closest old phase, that is metastable phase 1(MS1) from melt, MS2 from MS1,…, Stable solid (SS) phase from MS3 and each nucleation process is associated with a corresponding nucleation free energy as a function of nucleus size (R)(shown in **Figure 1(b)**). If the nucleation free energy curves of MS1 from melt and MS2 from MS1 cross each other at some nucleus size, R', we can write

$$\frac{4\pi}{3}R'^3 \Delta G_{Melt/MS1} + 4\pi R'^2 \gamma_{Melt/MS1} = \frac{4\pi}{3}R'^3 \Delta G_{MS1/MS2} + 4\pi R'^2 \gamma_{MS1/MS2}$$

$$R' = \frac{3(\gamma_{Melt/MS1} - \gamma_{MS1/MS2})}{(\Delta G_{MS1/MS2} - \Delta G_{Melt/MS1})} \quad (5)$$

This suggests that if free energy difference between Melt-MS1 is equal to MS1-MS2, phase MS2 cannot grow from MS1 in this energy diagram as $R' \to \infty$. Now, due to entropic contribution of all phases, curvatures of respective phases are different, that makes surface energy between two pairs also different. Generally, melt phase corresponds to the most flattened free energy surface (FES) and subsequent metastable phases with steeper FES. In this picture, to have a positive value of R' for sequential nucleation of MS1 in melt and MS2 in MS1, the condition is $\gamma_{MS1/MS2} > \gamma_{Melt/MS1}$ (due to different curvature) and $\Delta G_{MS1/MS2} < \Delta G_{Melt/MS1}$.

Next, an interesting situation arises for the nucleation of MS3 in MS2 which competes with the nucleation of MS1 in MS2. For the consecutive phase change from MS1 to MS2 followed by MS2 to MS3, if the nucleus size be R" where the two nucleation free energy curves cross each other (**Figure 1(b)**)

$$\frac{4\pi}{3}R'''^3\Delta G_{MS1/MS2} + 4\pi R''^2\gamma_{MS1/MS2} = -\frac{4\pi}{3}R'''^3\Delta G_{MS2/MS3} + 4\pi R''^2\gamma_{MS2/MS3}$$

$$R'' = \frac{3(\gamma_{MS2/MS3} - \gamma_{MS1/MS2})}{(\Delta G_{MS1/MS2} + \Delta G_{MS2/MS3})} \tag{6}$$

As by the analogy of curvature, $\gamma_{MS1/MS2} < \gamma_{MS2/MS3}$, nucleation of MS3 will be favored in MS2 (as shown in Figure 1(c)).

Now, with the difference in the free energy difference and the surface tension term, MS1 can also grow again in MS2 phase and system can be trapped in MS1. Figure 1(b) shows the free energy profiles required for subsequent nucleation where the values of radius of nucleus (R) is obtained by applying the analogy discussed above.

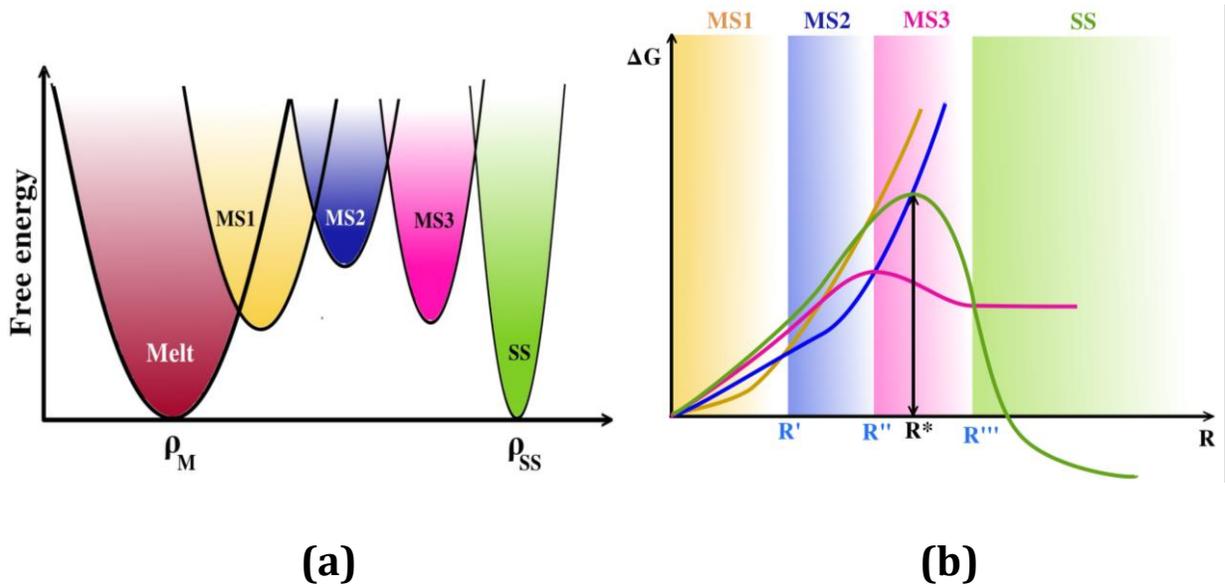

(a)  (b)

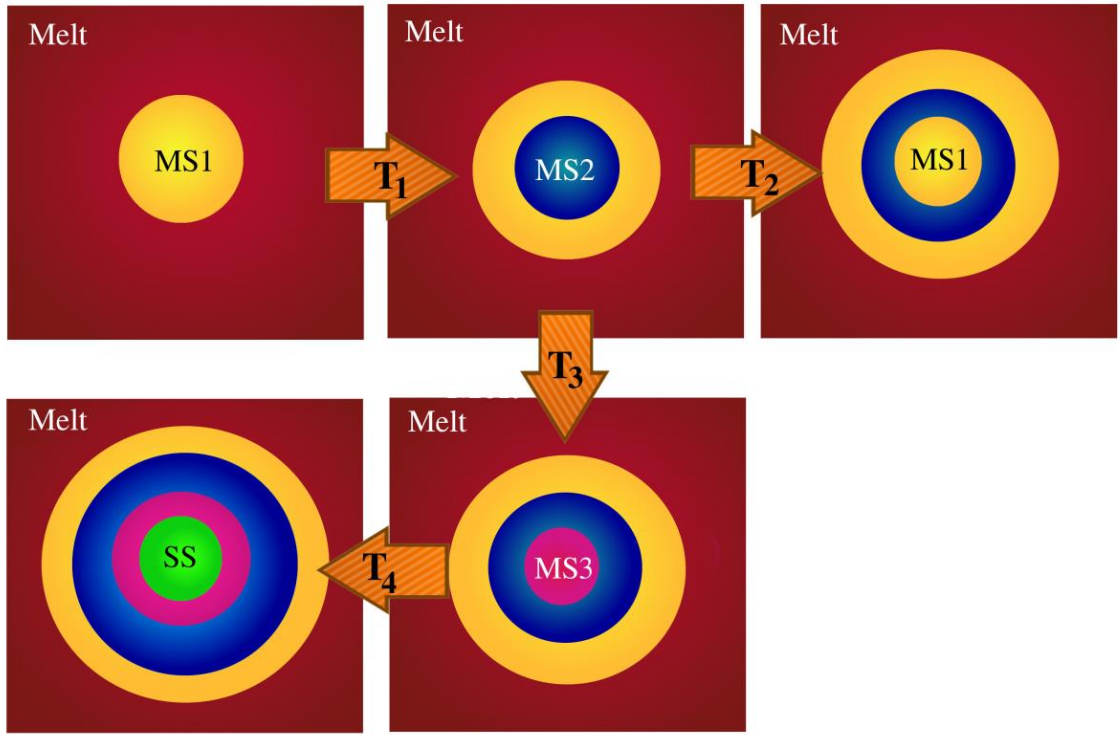

*Figure 1: A schematic representation of a growing nucleus wetted by less stable intermediate phases. (a) Free energy profile of two stable phases which are at coexistence in the presence of three metastable(MS) phases in a ladder like structure; all MS has higher energy than the stable phases. (b) Free energy of nucleation as a function of growing nucleus size at a constant temperature and pressure. Different size of the nucleus corresponds to lowest free energy of different phases. (c) Sequential nucleation of MS phases and stable solid phase from the melt phase as we increase temperature.*

We should emphasize that CNT provides perhaps the simplest possible explanation of kinetic control. However, as exemplified by OSR, the existing approaches of phase transformations based on CNT faces difficulty in explaining the occurrence of polymorphs and selection of phases. The inadequacy of CNT becomes apparent for the temperature dependent sequential phase transformation in zeolite or phosphate systems.

We can consider two systems with different energetics to understand the drawbacks of CNT (**Figure 2**). By a metastable solid, one usually considers a solid that is stable with respect to the melt (M) phase but metastable (MS) with respect to the final, more or most stable solid (SS) phase (**Figure 2(a)**). Here one compares the free energy barrier between M-MS and M-SS transformations. If the surface tension is not too disparate, and if the free energy of the SS phase is much lower than that of the MS phase, then the SS phase forms directly. However, if the surface tension of M-MS coexisting phases is much smaller than M-SS phases, then the MS phase might form first. If the solid remains in solution or in a mobile state, then this phase transformed to SS in due course of time. This is the Ostwald scenario.

Now, the interesting scenario comes when intermediate phase is slightly metastable with respect to the liquid phase itself (**Figure 2(b)**). In this case there is no barrier but there is a free energy cost of creating an embryo as the free energy is a monotonically increasing function of the radius R. The one dimensional nucleation free energy, $\Delta G(R)$ of CNT cannot describe the nucleation of MS from M in this case. This can be explained if we assume free energy of nucleation as a function of both R and an order parameter, Q (**Figure 2(c)**). That is, we adopt a two dimensional description for the free energy of the growing nucleus

$$\Delta G(Q,R) = \frac{4}{3}\pi R^3 \Delta G_v + 4\pi R^2 \gamma(R,Q) \tag{7}$$

Here, we assume that the order parameter Q can describe any conformational change of the system. Therefore, with respect to Q, MS state is associated with a minima in the two dimensional free energy surface. However, along R according to CNT free energy will be a monotonic increasing function upto the critical nucleus size for SS phase. Therefore, the nucleation through the MS phase can proceed along the saddle plane. The multidimensionality basically enters into the nucleation free energy by the dependence of surface tension on the order parameter, Q.

In this scenario, if the system gets trapped in the MS phase, it can give rise to another interesting situation. Depending on the nucleation barrier for (MS, M) and (MS, SS) it can either go to SS phase or come back to the original melt phase. An interesting class of polymorphism can be characterized as "Quickly Disappearing Polymorphs (QDPM)" [17]. In this case, the intermediate phase is so metastable that it disappears after forming and can come back again depending on the crystallization environment. J. Frenkel pointed out that the viscosity of the system increases in pre-freezing stage as the solid forms and melts [22]. One can argue that the quickly disappearing polymorphs are those states that are metastable with respect to both the liquid and the state solid (**Figure 2(b)**).

The nucleation barrier between (MS, M) and (MS, SS) depends largely on the order parameter value of MS phase ($\rho_{MS}$), here we have taken density as OP. We have carried out a simple calculation to see the dependence of lifetime of QDPM for the transitions of MS phase (phase 2) to original melt phase (phase 1) ($\tau_{12}$) or go the stable solid form (phase 3) ($\tau_{23}$). We can calculate surface tension, $\gamma_{12}$ and $\gamma_{23}$ from Cahn-Hilliard theory that we shall discuss in the next sections in detail ($\gamma_{12} = \sqrt{2\lambda\kappa}(\rho_1 - \rho_M)^2 + \sqrt{2\lambda\kappa}(\rho_1 - \rho_{MS})^2$). $\lambda$ parameter defines the curvature of the free energy surface of a particular phase, taken to be 1000 and $\kappa$ is related to the correlation length that is assumed to be $\lambda/2$. Here we have assumed the free energy difference between (MS and M), $\Delta G_{12}$ to be 10 and that between MS and SS, $\Delta G_{23}$ to be 20, $k_B T$ is considered to be 10. Now, we can calculate nucleation free energy barrier ($\Delta G^*$) from Eq. (2) and then calculate rate of nucleation and lifetime of the polymorph for two transitions using Arrhenius equation

$$k_{12} = A e^{-\Delta G_{12}^*/k_B T} \tag{8}$$

Now, if we change the OP for MS phase, $\rho_{MS,}$ the surface tension between two pair of phases will change that leads to the change in rate of transitions, $k_{12}$, $k_{23}$ and their lifetime, $\tau_{12}$ and $\tau_{23}$. In **Figure 2(d)**, we have shown the ratio of lifetimes as a function of the change in $\rho_{MS}$. When $\rho_{MS}$ is close to $\rho_{SS}$, the lifetime of transition MS to M ($\tau_{12}$) is large compared to $\tau_{23}$ as the surface tension between MS and SS is smaller. Therefore, MS goes to SS directly. However, as $\rho_{MS}$ comes close to $\rho_M$, $\tau_{12}$ becomes smaller and the MS state can come back to the original melt phase, that characterizes the disappearing polymorphs.

It is to be noted here that we have used the order parameter density to specify the phases. As mentioned earlier, change in that order parameter changes surface tension. However, we have to go back to one dimensional picture of CNT to calculate nucleation free energy.

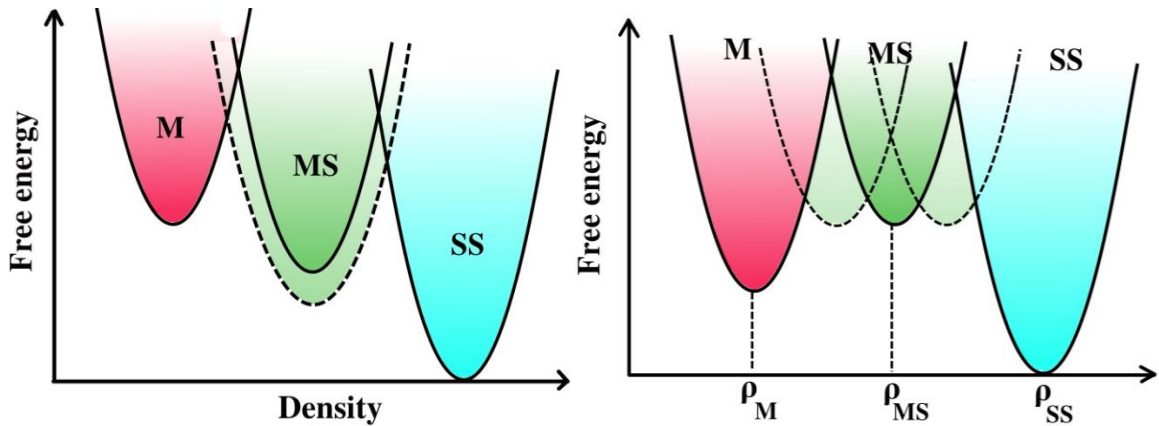

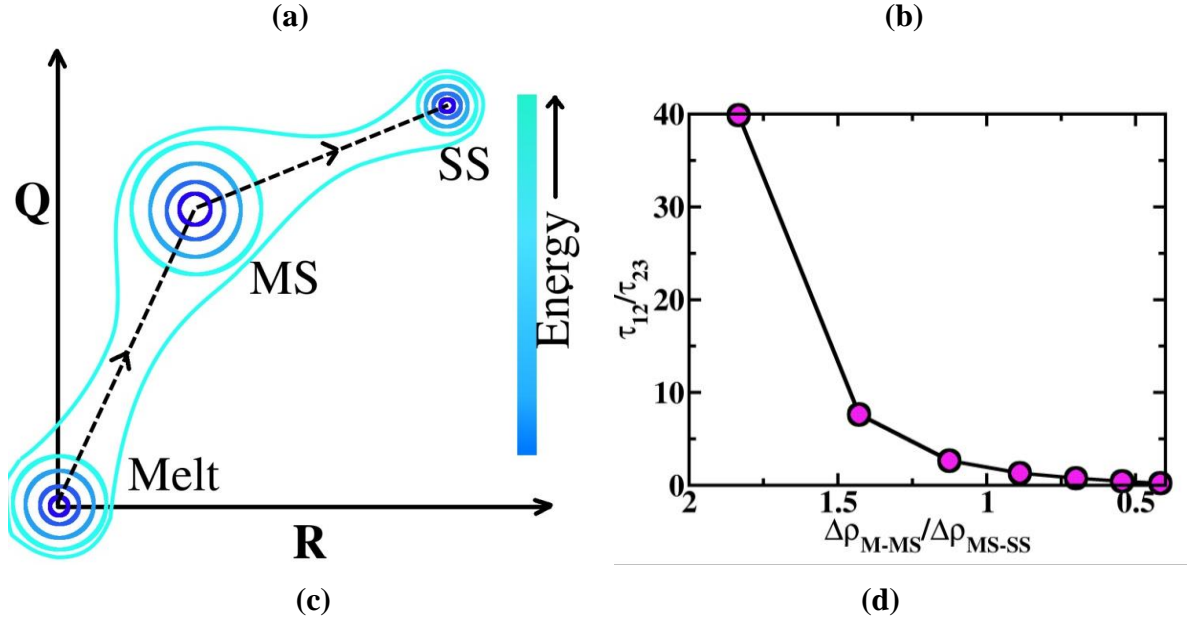

*Figure 2: (a) Free energy surface of melt and stable solid phase in the presence of one intermediate phase that is metastable with respect to stable solid phase, but stable with respect to the melt phase, (b) Free energy surface of melt and stable solid phase in presence of one intermediate phase that is metastable with respect to both melt and stable solid phase, (c) Free energy surface of nucleation as a function of two order parameters, radius of nucleus, R and a generalized order parameter for conformational change, Q, (d) Lifetime of the QDPM for the transitions from MS to melt ($\tau_{12}$) and MS to SS ($\tau_{23}$) as a function of the change in MS phase order parameter (here, density $\rho_{MS}$). When $\rho_{MS}$ is closer to $\rho_{SS}$, x axis value is greater than 1 and the transition MS to SS is more feasible ($\tau_{23} < \tau_{12}$) and when $\rho_{MS}$ is close to $\rho_M$, MS phase is more probable to come back to the melt.*

The ultimate selection may be guided by the domination of either the enthalpy or the entropy change. A consideration of the enthalpy-energy balance for a wide variety of metastable phases (diamond/graphite, various forms of silica, titania, various superconducting cuprates and microporous solids) leads to a natural classification of metastable solid phases into three categories : Type I (where change in enthalpy on phase separation, $\Delta H$, dominates), Type II (where change in entropy, $\Delta S$, dominates) and Type III (where both $\Delta H$ and $\Delta S$ contribute in comparable measure). For the latter, entropy arising from the vibrational degrees of freedom may play a critical role.

In a series of papers, Navrotsky and coworkers carried out calorimetric studies to investigate the phase transition of a number of important environmental and industrial minerals[8, 18, 23-25]. She used the thermodynamic data to investigate the selective polymorph formation at a particular temperature, pressure and composition of the mixture. In case of zeolite, she demonstrated that many metastable zeolite frameworks possess similar energy and entropy[26, 27]. We shall discuss this later in detail.

As already mentioned, the free energy barrier between two phases is decided partly by the surface free energy which is reduced in the presence of intermediate metastable phases. In a macroscopic picture, the dependence of surface tension on the micro droplet size is measured in terms of Tolman's length. This however gives a weak dependence. Using a completely different approach, Kirkpatrick-Thirumalai-Wolynes (KTW)[2] and Xia-Wolynes (XW)[3] proposed that surface tension varies with the radius of liquid droplet (r) following the relation

$$\gamma(r) = \gamma_0 \left(\frac{r_0}{r}\right)^{1/2} \tag{9}$$

Therefore, when the surface of the growing phase is coated with other intermediate phases, surface energy decreases, which is also predicted by random field Ising model [4]. Here we have formulated some model systems to further investigate the decrease of surface tension on the number of metastable phases by modeling the interface with the principle of having minimum surface energy where the more stable phases are coated with subsequent less stable phases.

No existing theoretical model can describe the nucleation of a polymorphic solids because of a number of difficulties. As temperature is raised both the curvature and the energy minima of the stable phases and intermediate phases change which leads to the crystallisation of most stable crystalline form. These aspects are very difficult to include in a theoretical model. Cahn-Hilliard (C-H) theory offers an expression of surface free energy for nonuniform systems. We shall discuss this in the next section in detail.

## III. Cahn-Hilliard theory of surface tension in the absence of any metastable intermediates

Cahn-Hilliard offers a simple, albeit phenomenological, approach to calculate the interfacial surface free energy of a non-uniform system [28]. It uses Ginzburg-Landau free energy functional to include the inhomogeneity of the two phases at the interface using a square gradient term of density fluctuation [29]. However, Cahn-Hillard theory, although phenomenological, allows analytical solution. In this approach the free energy of the system can be written as

$$\Omega_i[\rho(\mathrm{r})] = \int d\mathbf{r} \left[ f_{i,0}[\rho(\mathbf{r})] + \kappa \left( \nabla \rho(\mathbf{r}) \right)^2 \right] \tag{10}$$

Here $f_{i,0}[\rho(\mathrm{r})]$ is Landau (or, Helmholtz) free energy density function of the number density, $\rho(\mathrm{r})$ of the $i^{th}$ phase. $\kappa$ is related to the correlation length.

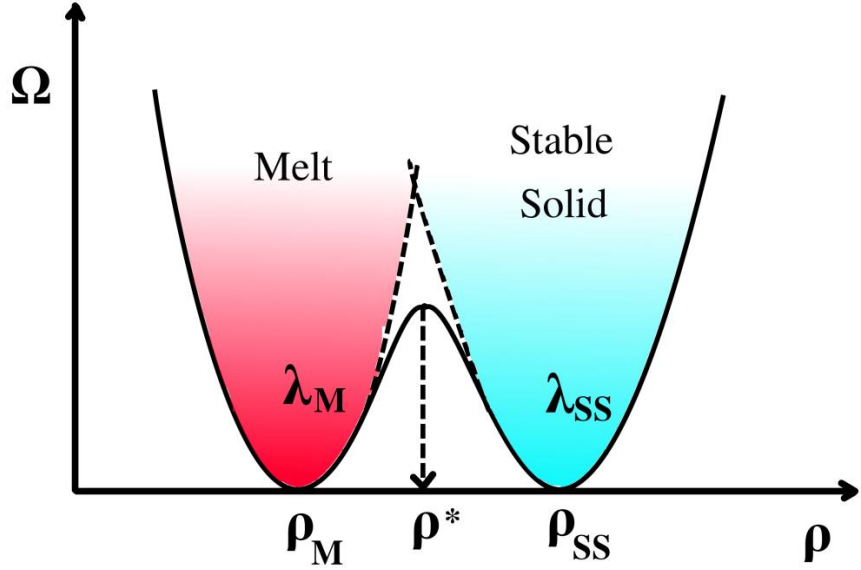

*Figure 3: Schematic interface profile of melt and stable solid at coexistence. Parameters of Cahn-Hilliard theory are shown: $\lambda_i$ are the curvatures of the respective parabolas, $\rho_i$ are the corresponding structural order parameters (density) of the two phases and $\rho^*$ is the point of intersection of two phases.*

If we take a flat interface at xy-plane, the problem becomes one dimensional and one can write the free energy by integrating over z.

$$\Omega_i[\rho] = A \int_{-\infty}^{\infty} dz \left[ f_{i,0}[\rho] + \kappa \left( d\rho/dz \right)^2 \right] \qquad (11)$$

Now, surface tension is the excess free energy per unit area ($\gamma = \Omega[\rho]/A$). Here, "A" is the area of the interface.

To define surface tension between two coexisting phases uniquely, it has to satisfy two criteria: i) the chemical potential of a particular species in the two phases should be same and ii) the thermodynamic grand potential density, $\omega_i$ of two phases should be same.

$$\mu_M(\rho_M) = \mu_{SS}(\rho_{SS})$$
$$\omega_M(\rho_M) = \omega_{SS}(\rho_{SS}) \, or, \, f_M - \mu_M \rho_M = f_{SS} - \mu_{SS} \rho_{SS} \tag{12}$$

By obtaining the free energy profile that minimizes surface energy and changing the integration variable from z to ρ(density), one obtains the surface tension

$$\gamma = 2 \int_{\rho_M}^{\rho_{SS}} [\kappa \Omega[\rho]]^{\frac{1}{2}} d\rho \tag{13}$$

If we assume piecewise harmonic potentials, $\lambda_M$ and $\lambda_{SS}$ for the two stable phases (melt phase (M) and stable solid phase (SS)) (shown in **Figure 3**), the grand potential density difference (relative to the bulk initial phase) can be written as

$$\Delta\omega_M = \frac{\lambda_M}{2}(\rho^* - \rho_M)^2$$
$$\Delta\omega_{SS} = \frac{\lambda_{SS}}{2}(\rho^* - \rho_{SS})^2 \tag{14}$$

Now, the total interfacial free energy can be divided into two parts, $\gamma_{total} = \gamma_1 + \gamma_2$ and written in terms of curvatures, $\lambda_i$ and square gradient coefficient, $\kappa$

$$\gamma_{Total} = \sqrt{2\lambda_M \kappa}(\rho^* - \rho_M)^2 + \sqrt{2\lambda_{SS} \kappa}(\rho^* - \rho_{SS})^2 \tag{15}$$

Now, in the presence of metastable phases between the stable solid and melt phase, the surface energy between two phases become modified. Here we will use a similar description of the system as of Granasy and Oxtoby[30](shown below), but with multiple metastable phases. Here we show the calculation of surface free energy in presence of three metastable phases for two different models: i) energy minima of all the metastable phases are equal, ii) energy minima of metastable phases have ladder like structure having highest energy for the middle most metastable phase(for odd number of MS)/phases(for even number of MS). However, in our calculation, we have varied the number of metastable phases from 0 to 10.

## IV. Surface Tension in the Presence of Multiple Intermediate Phases

We now discuss the system of main interest where the free energy surface is such that the minima of the two coexisting equilibrium phases is accompanied by a number of intermediate phases. As already discussed, the calculation of surface tension between two phases requires the coexistence between them. Now, if we consider all the intermediate phases and the stable phases are at coexistence and the curvatures of all free energy surfaces are equal, the surface tension between two stable phases (melt and stable solid) wetted by intermediate metastable phases ( $\gamma^{w}_{M/SS}$ ) is found to be related to the same without wetting ($\gamma_{M/SS}$) by the following relation (using Eq. (15))(full derivation is given in the *APPENDIX*).

$$\gamma^{w}_{M/SS} = \frac{\gamma_{M/SS}}{N+1}$$

(16)

where N is the number of metastable phases. We want to point out that there was a mistake in Eq. 10 of our previous paper[6] for this relation. Eq.16 (above) gives the correct expression.

However, this cannot be the real picture of nucleation as metastable phases are not expected to be at coexistence with the stable phases at a particular temperature and pressure. Now, we have taken two model systems where free energies of the metastable phases are higher than the stable phases but the models are distinct by the geometry and structures of free energy minima. It was discussed earlier that the presence of such intermediate metastable phases can profoundly affect the surface tension of the two coexisting phases.

## A. Model I

In this model system, all the metastable phases have same energy minima which is higher than the melt phase and stable solid phase which are at coexistence (**Figure 4**). The grand potential density differences of melt phase (M), stable solid phase (SS) and three metastable phases (MS$i$; $i$=1,2,3) relative to the bulk initial phases are given by

$$\Delta\omega_M = \frac{1}{2}\lambda_M(\rho-\rho_M)^2$$
$$\Delta\omega_{MSi} = \frac{1}{2}\lambda_{MSi}(\rho-\rho_{MSi})^2 + E_0; \quad i=1,2,3$$
$$\Delta\omega_{SS} = \frac{1}{2}\lambda_{SS}(\rho-\rho_{SS})^2$$

(17)

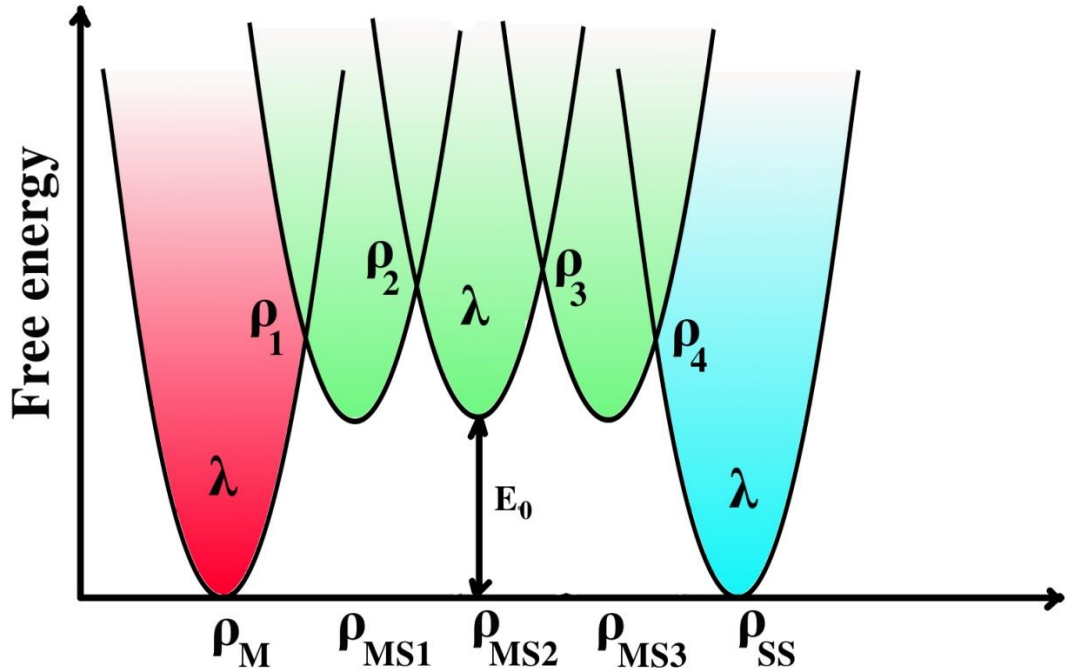

*Figure 4: Model system of interface between stable solid and melt phase with three metastable phases having same energy. Here the curvatures of all the free energy surfaces are assumed to be equal. Two stable phases are taken to be at coexistence to calculate surface tension between them.*

In this system, the total surface energy can be divided into eight parts. Now, if we assume the curvature of free energy surface ($\lambda_i$) for all the phases are equal, the total surface energy can be written as

$$\gamma = \gamma_1 + \gamma_2 + \gamma_3 + \gamma_4 + \gamma_5 + \gamma_6 + \gamma_7 + \gamma_8$$

$$= 2\sqrt{\kappa} \left[ \begin{array}{l} \sqrt{\dfrac{\lambda}{2}} \displaystyle\int_{\rho_M}^{\rho_1} (\rho - \rho_M) d\rho + \displaystyle\int_{\rho_1}^{\rho_{MS1}} \left[ \dfrac{\lambda}{2}(\rho - \rho_{MS1})^2 + E_0 \right]^{1/2} d\rho + \displaystyle\int_{\rho_{MS1}}^{\rho_2} \left[ \dfrac{\lambda}{2}(\rho - \rho_{MS1})^2 + E_0 \right]^{1/2} d\rho \\[2ex] + \displaystyle\int_{\rho_2}^{\rho_{MS2}} \left[ \dfrac{\lambda}{2}(\rho - \rho_{MS2})^2 + E_0 \right]^{1/2} d\rho + \displaystyle\int_{\rho_{MS2}}^{\rho_3} \left[ \dfrac{\lambda}{2}(\rho - \rho_{MS2})^2 + E_0 \right]^{1/2} d\rho + \displaystyle\int_{\rho_3}^{\rho_{MS3}} \left[ \dfrac{\lambda}{2}(\rho - \rho_{MS3})^2 + E_0 \right]^{1/2} d\rho \\[2ex] + \displaystyle\int_{\rho_{MS3}}^{\rho_4} \left[ \dfrac{\lambda}{2}(\rho - \rho_{MS3})^2 + E_0 \right]^{1/2} d\rho + \sqrt{\dfrac{\lambda}{2}} \displaystyle\int_{\rho_4}^{\rho_{SS}} (\rho - \rho_{SS}) d\rho \end{array} \right]$$

(18)

For the two terms ($\gamma_1$ and $\gamma_8$), with $E_0=0$, the solutions are

$$\gamma_1 = \sqrt{2\lambda\kappa}(\rho_1 - \rho_M)^2$$
$$\gamma_8 = \sqrt{2\lambda\kappa}(\rho_4 - \rho_{SS})^2$$

(19)

However, for the other terms it leads to the solution

$$\gamma_i = 2\sqrt{\kappa} \int_{\rho_i}^{\rho_{MSi}} \left[ \dfrac{\lambda}{2}(\rho - \rho_{MSi})^2 + E_0 \right]^{1/2} d\rho$$

$$= \sqrt{\dfrac{2\kappa}{\lambda}} E_0 \left[ \begin{array}{l} -\sqrt{1 + \dfrac{\lambda}{2E_0}(\rho_i - \rho_{MSi})^2} \left( \sqrt{\dfrac{\lambda}{2E_0}}(\rho_i - \rho_{MSi}) \right) \\[2ex] -\operatorname{arcsinh}\left( \sqrt{\dfrac{\lambda}{2E_0}}(\rho_i - \rho_{MSi}) \right) \end{array} \right]$$

(20)

## B. Model II

In the previous model, the energy minima of all the metastable phases are assumed to be equal which is unlikely for real systems. Therefore, here we include the modification that the metastable states have different energy minima which are arranged in a ladder like structure. However, all the free energy surfaces are assumed to have same curvature ($\lambda_i$). Now, for the system shown in **Figure 5**, with three metastable phases, the grand potential density differences of melt phase (M), stable solid phase (SS) and three metastable phases (MS$i$; $i=1,2,3$) relative to the bulk initial phase are given by

$$\Delta\omega_M = \frac{1}{2}\lambda_M(\rho-\rho_M)^2$$

$$\Delta\omega_{MS1/MS3} = \frac{1}{2}\lambda_{MS1/MS3}(\rho-\rho_{MS1/MS3})^2 + \frac{E_0}{2}$$

$$\Delta\omega_{MS2} = \frac{1}{2}\lambda_{MS2}(\rho-\rho_{MS2})^2 + E_0$$

$$\Delta\omega_{SS} = \frac{1}{2}\lambda_{SS}(\rho-\rho_{SS})^2 \qquad (21)$$

The final expression of total surface energy is similar given by Eq.(18), (19) and (20) but depending on the energy minima of the free energy surfaces ($E_0$ or $E_0/2$), the total surface energy changes.

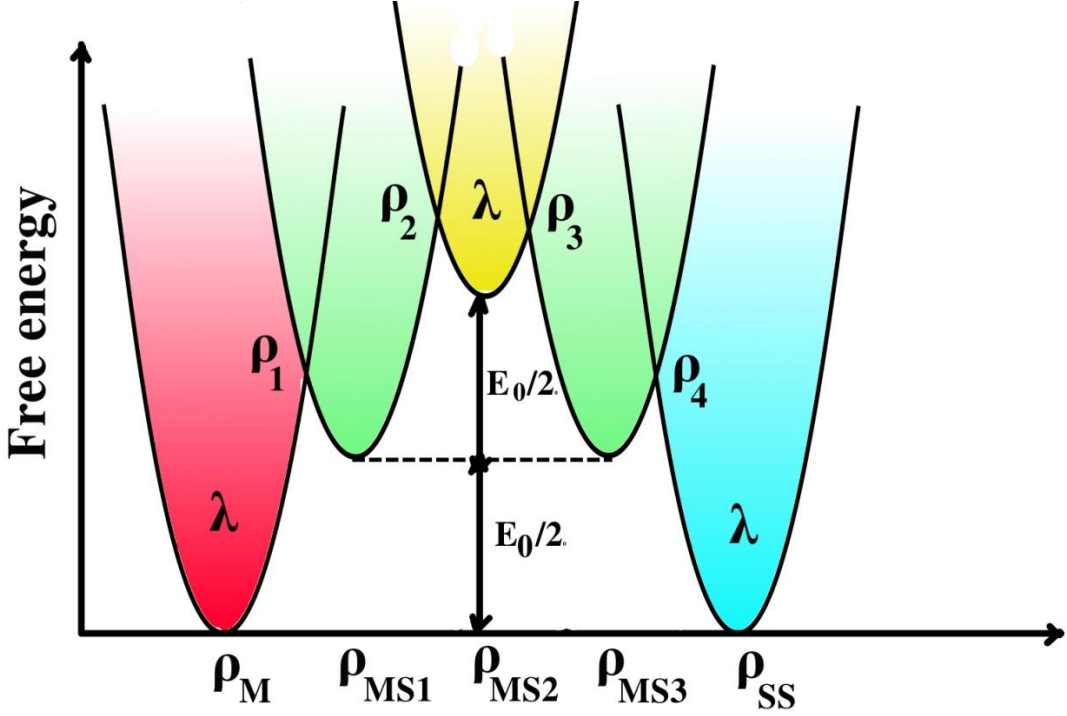

*Figure 5: Model system of interface between stable solid and melt phase with three metastable phases having different energy in a ladder like structure. However, the highest energy of the MS phase in this model($E_0$) is equal to the energy of MS phases in model I. The curvatures of the free energy surfaces are assumed to be equal. Two stable phases are taken to be at coexistence to calculate surface tension between them.*

In all the cases, it is assumed that all the free energy surfaces are equi-spaced. $\rho_M$ and $\rho_{SS}$ are the equilibrium densities of the melt and stable solid phase, respectively. $\rho_{MSi}$ is the equilibrium density of intermediate ith phase that is given as

$$\rho_{MSi} = \rho_M + i\Delta\rho$$
$$\Delta\rho = (\rho_{SS} - \rho_M)/(N+1) \tag{22}$$

where N is the number of metastable phases between melt(M) and stable solid (SS) phase.

## V. Numerical results and discussions

As already mentioned in the Introduction, it is notoriously difficult to obtained values of surface tension between any two phases, because of the condition of coexistence. In addition, reliable values of the free energy difference are also meager. Therefore, we have assumed certain reasonable values of these parameters and carried out calculations which are demonstrative of the possibilities as predicted by the theory.

Here we compute total surface free energy between melt (M) and stable solid (SS) phases in presence of multiple metastable phases ranging from 0 to 10. The value of the parameters we used in our calculations are: curvature of the free energy surface ($\lambda_i$) =3000, $\kappa = \lambda/2$, $\rho_M$=0.88, $\rho_{SS}$=1.05.

We have used a set of $E_0$ values from 0.1 to 5.0 (which is less than the energy where free energy surfaces of melt and stable solid surfaces cross each other, 10.837). **Figure 6** depicts surface free energy of stable melt and crystalline solid phases in presence of multiple metastable phases with different $E_0$ values for model I and model II.

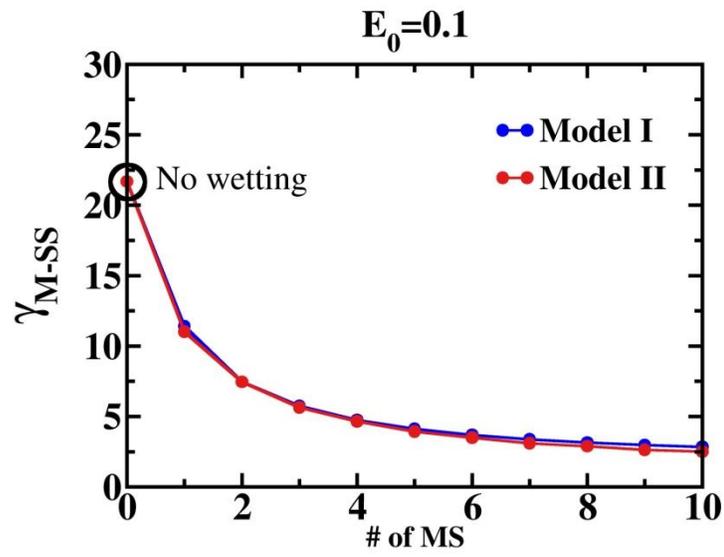

(a)

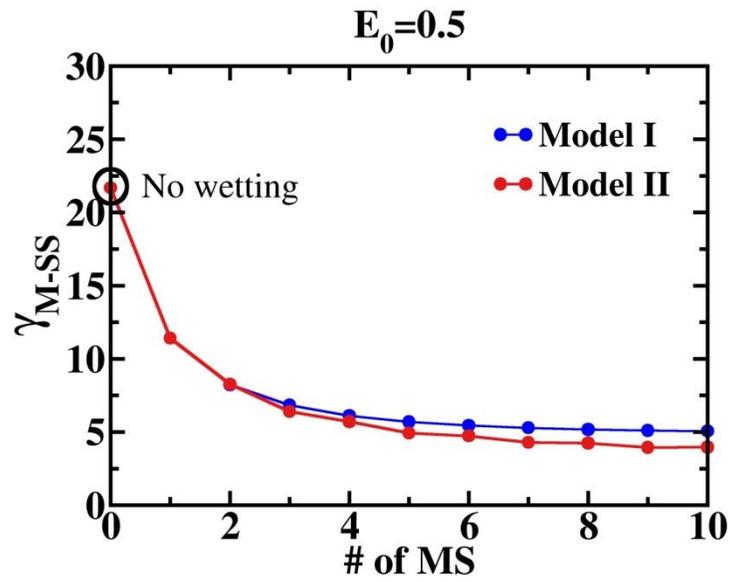

(b)

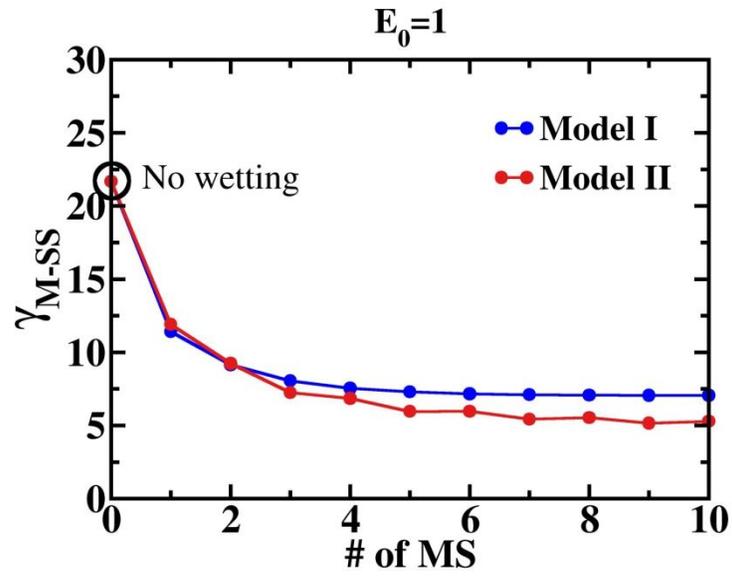

**(c)**

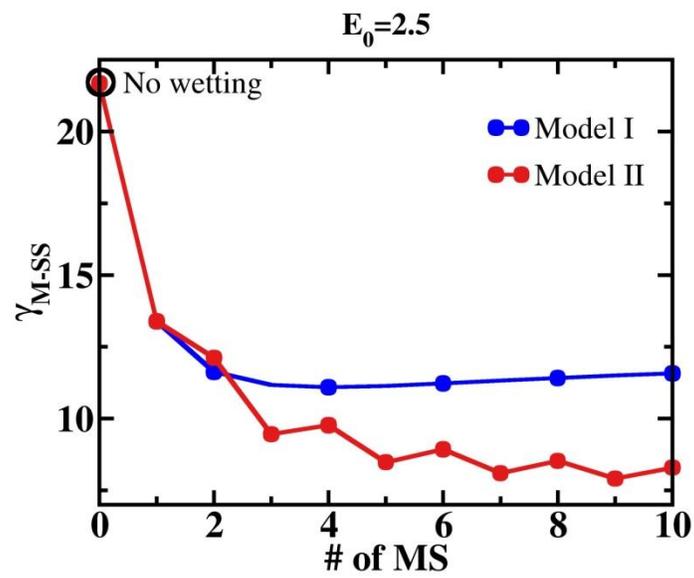

**(d)**

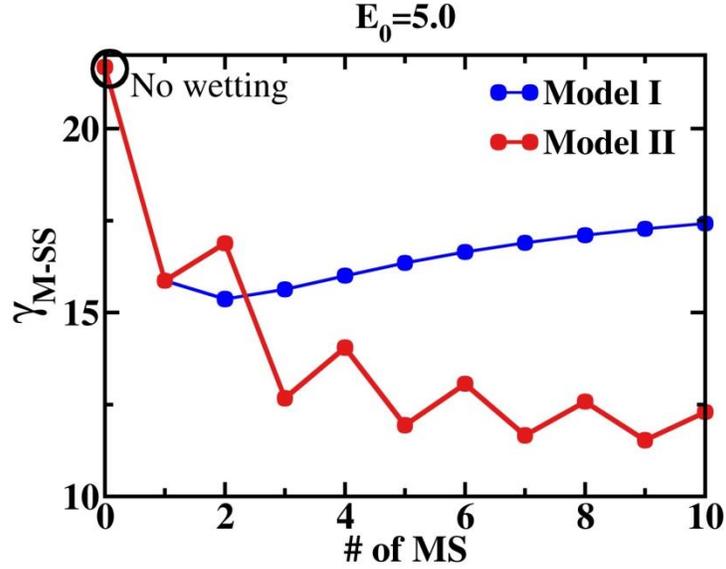

(e)

*Figure 6: Surface free energy of melt and stable solid phase in presence of multiple metastable phases. (a)-(e) correspond to different energies ($E_0$) of metastable phases for model I and model II. At the higher $E_0$ values ($E_0$= 2.5 and 5.0), surface energies of model II with different number(N) of MS shows a zigzag pattern because of the structure of free energy surfaces in model II. When we add even number of MS phase, keeping $E_0$ fixed, surface energy does not decrease, rather increases. But when we add an extra odd number of MS phase (keeping $E_0$ fixed) all the FES become more compact and surface energy decreases.*

It is evident from **Figure 6** that when the energy of metastable phases is sufficiently lower and closer to two stable states (M and SS) (**Figure 6((a), (b) and (c))**), surface tension between two stable states ($\gamma_{M-SS}$) is reduced significantly in both the models. However, for a moderate $E_0$ value like 2.5 (**Figure 6(d)**), $\gamma_{M-SS}$ of Model II become significantly lower than Model I with increase of the number of MS.

However, at the value of $E_0$ equal to 5.0(**Figure 6(e)**), $\gamma_{M-SS}$ for Model II is reduced as number of metastable states increases and become almost half of the no wetting condition surface tension when number of MS states is 10, but $\gamma_{M-SS}$ for Model I decreases initially with increasing number of MS and it has a minima at 2. But after that it increases as number of MS become higher (although it is lower than no wetting condition). This is the artifact of model I system. As we pack more and more metastable states between two stable states at the same $E_0$ value (significantly higher at 5.0), the situation become more unreal. On the other hand, in a ladder-like structure in model II, surface tension decreases significantly in stead of a high free energy of metastable states.

As already discussed, we have derived an analytical expression for the surface tension between two stable phases in the presence of N number of intermediate metastable phases (Eq. (16)). However, this relation is obeyed only if all the metastable phases are considered to be at coexistence with the two stable phases. In models I and II, there are two different arrangements of free energy minima of metastable phases and the dependence of surface tension on number of metastable phases (N) gets modified. We shall address this issue with an asymptotic analysis in our future works and may connect these results with those from spin glass theories [2,4].

## VI. Role of temperature on phase transformation

We have already mentioned in the introduction that in case of polymorphic systems, at different temperatures different forms crystallize out. Here we want to analyze the effect of temperature on the surface tension between melt and stable solid phase in the presence of metastable phases. We have assumed that

i) At low temperature the free energy surfaces of metastable phases become steeper (with higher curvatures) and as temperature increases the curvatures become lower. Also, we have assumed here that the change in curvature of the melt and stable solid phases are negligible compared to the metastable phases.

ii) The relative stability of melt phase, metastable phases and stable solid phase (coexisting with the melt) remain constant with increasing temperature.

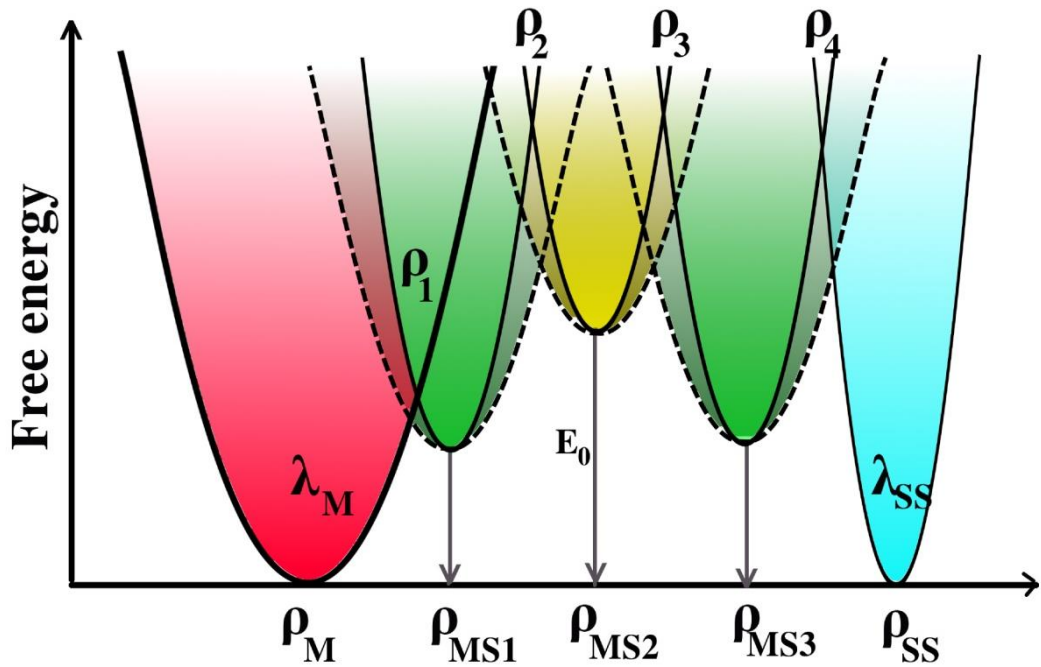

(a)

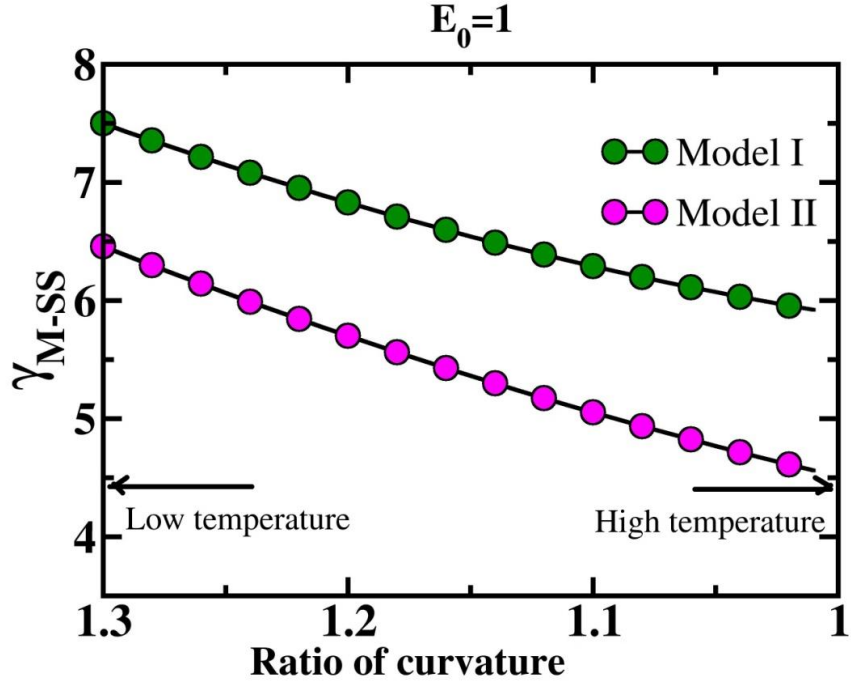

**(b)**

*Figure 7: (a) Free energy surface of melt phase (M), stable solid (SS) and metastable phases (MS) with different curvatures ($\lambda_i$) for a model II type system (although we have carried out numerical calculation of the change in surface tension with curvature of the FES change for both the model systems. We have changed the ratio of curvatures of metastable phases with the bulk phase. (b) Surface tension between melt and stable solid phase when the curvature of the MS is modified. Here the ratio of curvature is in between melt phase and MS phases.*

**Figure 7(a)** shows the geometry of the system where melt and stable solid phase have different curvatures and all the MS phases have same curvature which is varied in a range to model the change in the system energetic induced by change in temperature. Initially, we have taken the curvature of the melt phase, MS states and stable solid states to be 1500, 2000 and 3000 respectively. Then we have gradually changed the curvatures of the MS phases from 2000 to 1500 to model the effect of increasing temperature. We have shown the surface tension between melt and stable solid phase using model I and model II as a function of the ratio of the curvatures between MS phases and melt phase in **Figure 7(b)**. It is clear from the figure, when temperature increase (ratio of curvature goes towards 1), surface tension between melt and stable solid decreases significantly for both model systems with $E_0=1.0$. Therefore, in the high temperature the stable solid is crystallized out from the melt phase.

## VII.     Applications to real polymorphic systems : Zeolites and Phosphates

Many important minerals are polymorphic in nature. Ostwald step rule (OSR) has been applied repeatedly to understand sequential nucleation of these polymorphic phases, but no quantitative theoretical description is available to study the nucleation of such systems. In practical applications, generally a specific polymorph is precipitated out by selecting a particular temperature and pressure. A theoretical model to describe this polymorph selection faces enormous difficulty because with the change in temperature and pressure a number of parameter changes in the free energy surface of the system: the free energy minima, curvature of the free energy surfaces etc. We have already computed surface tension between two stable phases in presence of multiple metastable phases by varying these parameters in the last section. Now, we shall try to describe some real polymorphic systems which can be modeled by the two model systems we have considered in this work.

## A. Zeolites

As discussed earlier, Navrotsky and coworkers studied thermochemical properties of zeolite polymorphs such as FAU, MFI, BEA, FER etc. having different pore size, accessible surface area, largest cavity size etc. They showed that many zeolite frameworks are similar in energy and entropy and only slightly higher in energy than the stable polymorph, quartz [26, 27]. Therefore, the synthesis of a particular framework requires not only a particular temperature and pressure but also a structure directing agent typically alkylammonium, nitrogen containing organic molecules etc.

However, in an experimental study of adiabatic calorimetry, Wang *et al.* [31] observed that the surface entropy has a minor contribution (TΔS) to the total free energy of nanocrystalline phase of CoO (~1.5 J K$^{-1}$ mol$^{-1}$). It has been found true for other systems also. Therefore, the total surface free energy of polymorphic phases generally follows the pattern of surface enthalpy.

An accurate calculation of surface entropy is prohibitively difficult both by experiment and theory. Analytically, the vibrational entropy corresponds to a harmonic oscillator frequency of ω, can be calculated using the following well-known expression[5]:

$$S_v = R \left[ \frac{\hbar\omega}{k_B T \left(e^{\hbar\omega/k_B T} - 1\right)} - \ln\left(1 - e^{-\hbar\omega/k_B T}\right) \right] \tag{23}$$

Here $k_B$ is the Boltzmann constant, h is Planck's constant, R is the universal gas constant and T is the absolute temperature of the system. However, as the frequency of the harmonic surface approaches zero limit, the contribution of soft vibrational modes to entropy can be replaced by the corresponding rotational modes (using Rigid-Rotor-Harmonic-Oscillator (RRHO) Approximation). Grimme studied thermodynamics of supramolecular assembly by dispersion corrected density functional theory (DFT) and proposed an equation for vibrational entropy for the low frequency modes [32]

$$S(\omega) = w(\omega)S_V + [1 - w(\omega)]S_R \qquad (24)$$

where $S_V$ is the vibrational entropy and $S_R$ is the rotational entropy. The total entropy interpolates between $S_V$ at higher $\omega$ and $S_R$ at lower $\omega$ with the help of the Head-Gordon damping function [33], $w(\omega)$ (with a cut off frequency $\omega_0$), that is defined as

$$w(\omega) = \frac{1}{1 + (\omega_0/\omega)^a} \qquad (25)$$

In this case, we have shown a schematic free energy surface for stable phases and intermediate metastable states in **Figure 8** following the thermodynamics data available in literature[26]. Here we have taken two metastable framework structure, MFI and FER. They have different molar volume, density(shown in the figure) as well as poresize. But their energetic are similar. Both structures are associated with a transition enthalpy of 6-7 kJ/mol with respect to the most stable phase, quartz. ( $\Delta H_{trans}^{FER}$ =6.6 kJ/mol, $\Delta H_{trans}^{MFI}$ =6.8 kJ/mol at 298 K). These metastable structures are close to stable phase, quartz in both stability and density. Transition enthalpies of them are in the order of 2-3 $k_B T$. However, as we have discussed earlier, at low temperature the phase that separates from aluminosilicate melt is FAU (framework density= 13.45 Si/nm$^3$ and $\Delta H_{trans}^{FAU}$ =13.6 kJ/mol at 298 K). However, there are other phases like CHA with similar transition enthalpy but different density (framework density= 15.4 Si/nm$^3$ and $\Delta H_{trans}^{CHA}$=11.4 kJ/mol at 298 K). Both these structures posses similar but higher transition enthalpy (~4-6 k$_B$T) compared to MFI, FER. This types of metastable free energy arrangements are similar to our model system I.

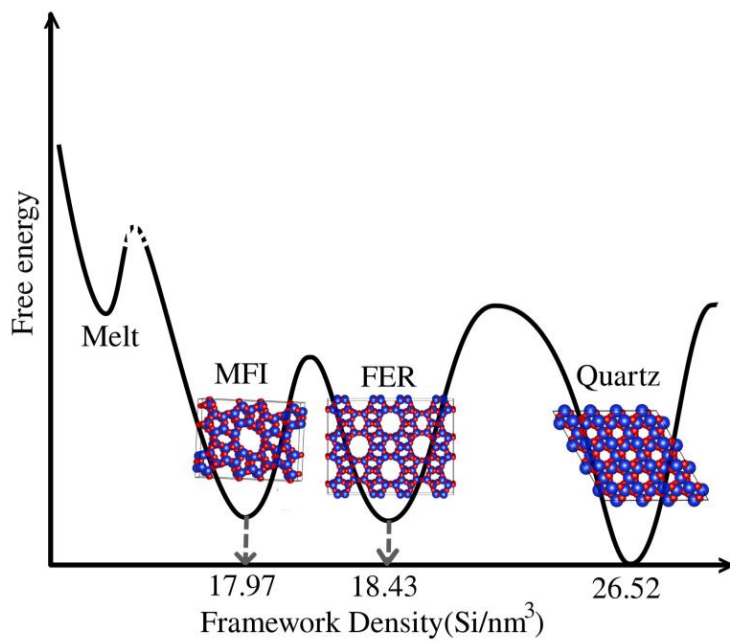

*Figure 8: Schematic free energy surface for zeolite polymorphs with most stable structure quartz and different metastable framework structures, MFI and FER having similar energy which is slightly higher than quartz (enthalpy of transition of these states to quartz is in the order of 2-3 k$_B$T [Ref [26]].*

## B. Phosphates

Hydroxyapetite(HAP), $Ca_{10}(PO_4)_6(OH)_2$ is one of the most important biomineralization process in living organisms, especially in vertebrates, as it is the primary mineral of hard tissues like bones and teeth [34]. The nucleation of this mineral has been a subject of many research works for last few decades and the process is found out to occur through an intermediate precursor phase, octacalcium phosphate (OCP), $Ca_8H_2(PO_4)_6.5H_2O$ [35]. Further, the nucleation from OCP to HAP has been observed to involve one hydrated surface layer of HAP [36]. We have shown a schematic free energy profile for this nucleation process in **Figure 9**.

As demonstrated in section IV, the presence of metastable phase reduces the surface free energy between OCP and HAP. Here the energy difference between the intermediate metastable phase and the stable HAP solid is significant (**Figure 9(a)**). Because hydration of HAP causes distortion of structure due to strong interaction between water molecules and phosphate and calcium. Therefore, according to classical nucleation theory (Eq.(1)), here free energy of nucleation contains both reduced (due to presence of intermediate state) positive contribution from surface tension term and negative contribution from free energy gap.

**Figure 9(b)** shows a schematic nature of free energy of nucleation ($\Delta G$) as a function of growing nucleus size (R) where the critical nucleus size of the two phases are different with different nucleation barrier. The fact that the metastable hydrated HAP phase is higher in energy but possess lower free energy barrier of nucleation than the stable HAP phase. This causes two free energy curves to cross each other and at the crossing point both the phases can exist in equilibrium with the bulk phase of octacalcium phosphate. This results in a "triple point" in the phase diagram.

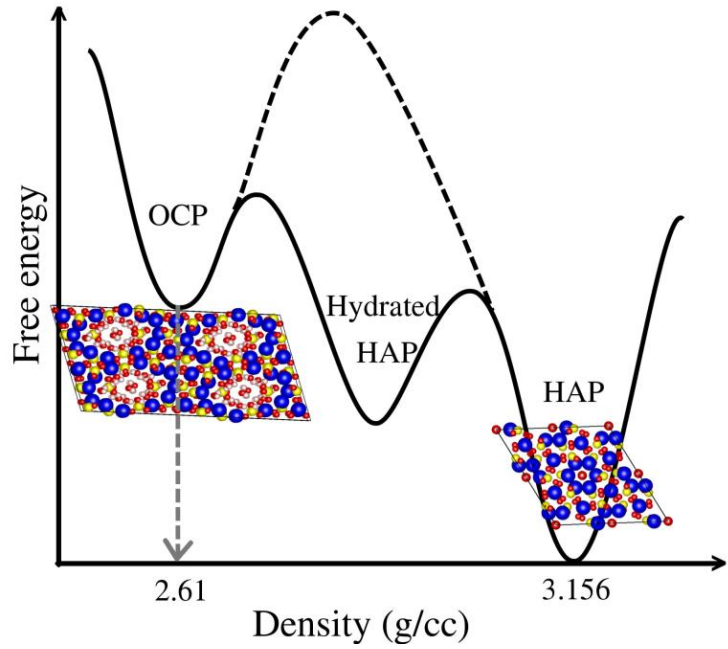

**(a)**

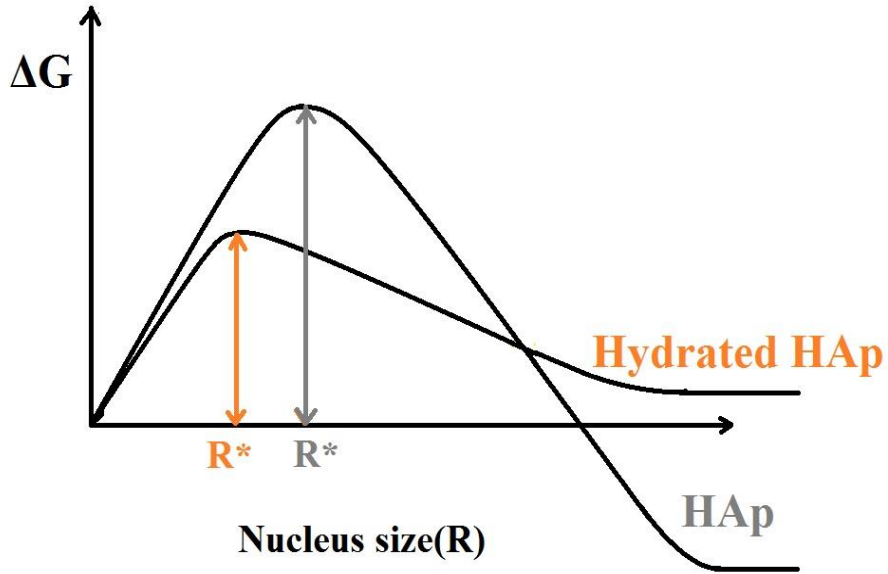

**(b)**

*Figure 9: (a) Schematic free energy surface of the octacalcium phosphate(OCP)-Hydroxyapetite(HAP) interface in the presence of metastable hydrated layer [Ref. [37] [38]]. (b) Free energy of nucleation as a function of nucleus size for metastable phase(hydrated HAP) and stable phase (HAP).*

## C. Energy materials

Another emerging field where polymorph control is crucial for organic small molecules that are used in organic light emitting diode (OLED), photovoltaic cells, memories, thin film transistors etc[31-32]. A slight change in molecular packing leads to a major change in interaction energy for different π-π stacking distance. Therefore, the metastable polymorphs having different packing possess a significant difference in their electronic properties and one can select a specific polymorph in these systems to attain high device performance. Similar to the systems described above here also depending on the free energy minima and the surface tension of metastable phases and the polymorph we want, nucleation free energy varies and a particular polymorph is selected in a specific temperature and pressure condition.

## VIII. Rugged multidimensional barrier surface

Due to the presence of the multiple metastable phases at intermediate values of the order parameter, the barrier surface could get modified significantly from what is envisaged in the CNT. In a one dimensional CNT description, the effects of the metastable phases are included through the surface tension, so long the value of the surface tension is taken from experimental measurements. Theoretical calculations can severely over-estimate the value of the surface tension between two coexisting (equal free energy) stable phases if the influence of the metastable phases is neglected. All these have been discussed above.

The one dimensional description may be inadequate to describe the transient trapping and return to the original phase (usually the melt). However, the transport coefficients like the diffusion coefficient could be quite different in those trapped states than the melt.

This raises an interesting yet difficult question. In most nucleation rate calculation of solid from the melt, the nucleation rate is assumed to be proportional to the diffusion coefficient of the melt. Anomalously slow nucleation and glassification is often attributed to the sharp decrease in the self-diffusion coefficient of the water molecules in the supercooled liquid state.

The present work broadens this perspective. The diffusion coefficient of the melt can become slow due to transient trapping – an aspect partly neglected in most contemporary studies. Therefore, at low temperatures, the probability of being trapped in one of the metastable states increases.

In the presence of a multidimensional surface, the nucleation pathway could be rather different from what is assumed in the CNT.

# IX. Conclusion

In this work, we first analyze how the presence of intermediate metastable phases could affect the surface tension between two stable phases: the melt and the stable solid phase which are at coexistence. We have used Cahn-Hilliard theory to calculate surface tension throughout this paper. At first we have taken two model systems with all the free energy surfaces of stable states and MS states to be equal but the free energy minima of the MS phases have different geometric arrangements (Model I and Model II). We have calculated surface tension in these two model systems with different value of the free energy minima of the MS phases ($E_0$) and obtained the dependence of surface tension on the number of metastable states as we change the energetics ($E_0$) and the geometry of the free energy minima of the system. We have also modeled the different polymorph selection with changing temperature by varying the curvature of MS phases in a simple model system. Computed surface tension between melt and stable solid phase in these model systems clearly describes the crystallisation of the stable crystalline form from the melt phase that becomes possible only at higher temperature. Finally, all these model systems are shown to describe a few real polymorphic systems like zeolite, hydroxyapetite and energy materials like organic electronics etc. However, there is a serious lack of accurate surface energy data for all these systems to decide the nucleation free energy. A quantitative simulation study can help in this context.

A major outcome of the present study is a multidimensional perspective of the phase separation kinetics in a complex solid characterized by many free energy minima. Phase separation in such a landscape cannot be described by a one dimensional treatment like CNT. Notable deficiency is the failure to describe the "quickly disappearing polymorphs" which are shown here to arise from trapping in metastable minima, a trapping facilitated by small values of surface tension due to the proximity of the metastable minima, in the order parameter plane, to that of the melt.

While the capillary approximation adopted in CNT allows to include the effects of multiple metastable minima on the experimentally observed surface tension, thus allows inclusion of these effects on the nucleation free energy surface, the failure of this approach becomes clear when we carry a density functional theoretic analysis, as was done in Ref. 6. The DFT calculation does not make the capillary approximation. Instead obtain the order parameter profile by directly minimizing the free energy with respect to the profile itself, and then obtains the free energy barrier. The result of such an analysis provides a detailed and rich picture of nucleation and growth which is different in the presence of metastable phases from that in the absence of those. The absolute value of the free energy barrier obtained from DFT is quite different that from CNT.

*The present work reveals a remarkable result that the surface tension of the two stable coexisting phases (for example, the melt and the solid phase) decreases as $1/(N+1)$ with the number $N$ of the metastable phases with intervening values of the order parameter(s).* This calculation assumes that the metastable phases are characterized by equally spaced equally deep free energy minima along the order parameter coordinate which are at coexistence with two stable phases. This is an asymptotic result because of the assumption of equal free energy of all the intermediate phases. This ($1/(N+1)$) term gets modified by the effect of the height of the

minima. We have presented an explicit calculation of the dependence of surface tension when the heights of the minima are different.

We should emphasize the semi-quantitative nature of the present work. While we have been able to provide a more advanced explanation of the effects of metastable phases on nucleation of solids with many polymorphs, the present analysis suffers from the limitation of non-availability of accurate free energy surfaces that would allow a more quantitative estimation of the nucleation rate. Such an analysis would require extensive work with specific systems, with force fields not yet fully available.

## Acknowledgement

The work was supported partly by Department of Science and Technology (DST), Govt. of India, Sir J. C. Bose fellowship, and Council of Scientific and Industrial Research (CSIR), India.

[36] J.-C. Heughebaert and G. Montel, Calcified tissue international **34**, S103 (1982).

[37] Y. Liu, R. Shelton and J. Barralet, presented at the Key Engineering Materials, 2004 (unpublished).

[38] J. Elliot, Structure and Chemistry of the Apatites and Other Calcium Orthophosphates (1994).

# APPENDIX A.

At first we consider the situation where no intermediate phase is there between two stable forms. Equation (15) gives the expression for surface tension in the system shown in Figure 3.

Now, we define density difference, $\quad \Delta\rho = \rho_{SS} - \rho_M \qquad (26)$

As we assume here, the curvature of two surfaces are equal, the crossing point of two surfaces can be defined as

$$\begin{aligned} \rho^* &= \frac{\rho_M + \rho_{SS}}{2} \\ &= \rho_M + \frac{\Delta\rho}{2} \end{aligned} \qquad (27)$$

$$\rho_{SS} = \rho_M + \Delta\rho \qquad (28)$$

Substituting Eq. (27) and (28) in Eq. (15), we get

$$\gamma_{M/SS} = \gamma_1 + \gamma_2 = 2 \times \frac{1}{2} \frac{\sqrt{2\kappa\lambda}}{4} (\Delta\rho)^2 = \frac{\sqrt{2\kappa\lambda}}{4} (\Delta\rho)^2 = \frac{\sqrt{2\kappa\lambda}}{4} (\rho_{SS} - \rho_M)^2 \qquad (29)$$

Now, if there are N number of metastable phases between M and SS, all of which are at coexistence with M and SS and all of them have same curvatures, the free energy density of different states can be written as

$$\Delta\omega_M = \frac{1}{2}\lambda(\rho-\rho_M)^2$$
$$\Delta\omega_i = \frac{1}{2}\lambda(\rho-\rho_i)^2 \quad (30)$$
$$\Delta\omega_{ss} = \frac{1}{2}\lambda(\rho-\rho_{SS})^2$$

where $\Delta\omega_i$ is for $i^{th}$ intermediate metastable state (i=1…N). We assume curvature, $\lambda$ is same for all.

Now, for 2 metastable states in between melt and solid phases,

$$\Delta\rho = \frac{\rho_{SS}-\rho_M}{2+1}$$
$$\rho_i = \rho_M + i\Delta\rho; i=1 \text{ for MS1}, i=2 \text{ for MS2}$$
$$\rho_1 = \frac{\rho_M+\rho_{MS1}}{2} = \rho_M + \frac{\Delta\rho}{2} \quad (31)$$
$$\rho_2 = \frac{\rho_{MS1}+\rho_{MS2}}{2} = \frac{\rho_M+\Delta\rho+\rho_M+2\Delta\rho}{2} = \rho_M + \frac{3}{2}\Delta\rho$$
$$\rho_3 = \frac{\rho_{MS2}+\rho_{SS}}{2} = \frac{\rho_M+2\Delta\rho+\rho_M+3\Delta\rho}{2} = \rho_M + \frac{5}{2}\Delta\rho$$

Here, the integral for surface free energy (Eq.(13)) can be divided into six parts:

$$\gamma = \gamma_1 + \gamma_2 + \gamma_3 + \gamma_4 + \gamma_5 + \gamma_6$$

$$= \sqrt{2\kappa\lambda} \left[ \begin{array}{l} \int_{\rho_M}^{\rho_1} (\rho - \rho_M) d\rho + \int_{\rho_1}^{\rho_{MS1}} (\rho_{MS1} - \rho) d\rho + \int_{\rho_{MS1}}^{\rho_2} (\rho - \rho_{MS1}) d\rho \\ + \int_{\rho_2}^{\rho_{MS2}} (\rho_{MS2} - \rho) d\rho + \int_{\rho_{MS2}}^{\rho_3} (\rho - \rho_{MS2}) d\rho + \int_{\rho_3}^{\rho_{SS}} (\rho_{SS} - \rho) d\rho \end{array} \right]$$

$$= \frac{1}{2}\sqrt{2\kappa\lambda}\left[(\rho_1 - \rho_M)^2 + (\rho_{MS1} - \rho_1)^2 + (\rho_2 - \rho_{MS1})^2 + (\rho_{MS2} - \rho_2)^2 + (\rho_3 - \rho_{MS2})^2 + (\rho_{SS} - \rho_3)^2\right]$$

$$= 6 \times \frac{1}{2}\sqrt{2\kappa\lambda}\left(\frac{\Delta\rho}{2}\right)^2$$

(32)

Now, using the expression of $\Delta\rho$ from Eq.(31), we get

$$\gamma_{M/SS}^w = \frac{1}{2}\sqrt{2\kappa\lambda}\frac{(\rho_{SS} - \rho_M)^2}{4} \times \frac{6}{(3)^2}$$

$$= \frac{1}{2}\sqrt{2\kappa\lambda}\frac{(\rho_{SS} - \rho_M)^2}{4} \times \frac{(2N+2)}{(N+1)^2} ; here\ N = 2\ for\ MS1\ and\ MS2 \qquad (33)$$

$$= \sqrt{2\kappa\lambda}\frac{(\rho_{SS} - \rho_M)^2}{4} \times \frac{1}{(N+1)}$$

The comparison of Eq. (33) with Eq. (29) gives Eq. (16). Please note that we gave a wrong equation ( Eq. 10) for this relation in our previous paper[6].